\documentclass[epj]{svjour}

\newcommand{\ttbar}{t\bar{t}}

\newcommand{\pt}{p_{\rm T}}
\usepackage{graphics}
\begin{document}
\title{A study of top-quark mass measurement
using the lepton energy distribution at the Large Hadron Collider}

\author{Seo Hyun An\inst{1} \and Sayaka Kawabata\inst{2} \and Tae Jeong Kim\inst{1}\thanks{\emph{e-mail:} taekim@hanyang.ac.kr } % etc
}                     % Do not remove
\institute{
          Department of Physics, Hanyang University, Seoul 04764, South Korea\label{1}
          \and
          Institute of Convergence Fundamental Studies, Seoul National Universiy of Science and Technology, Seoul 01811, South Korea\label{2}
}
\date{Received: date / Revised version: date}
%\date{7 March 2017 }
\abstract{
We present a feasibility study of top-quark mass measurement using 
the lepton energy distribution in the top-quark decay $t \to bW \to b\ell\nu$ at the LHC.
The method used in this study requires only the lepton energy distribution at parton level.
The analysis is performed in the lepton + jets final state by using fast simulation
data corresponding to an integrated luminosity of approximately 20 fb$^{-1}$ at $\sqrt{s}$ = 14 TeV.
Events with exactly one lepton, at least four jets and two $b$ jets are selected.
The lepton energy distribution at parton level is obtained by applying the bin-by-bin unfolding technique.
The study shows that the pole mass of the top quark can be measured with an uncertainty of the order of 1 GeV.
 % end of PACS codes
} %end of abstract

\titlerunning{top-quark mass using lepton energy distribution}
\authorrunning{S. H. An \and S. Kawabata T. J. Kim} 

\maketitle

\section{Introduction}
\label{sec:intro}

The top quark, which has the largest mass in the Standard Model (SM), is believed to give us a hint for new physics
due to its inevitable large coupling with the Higgs boson discovered in 2012 at the Large Hadron Collider (LHC).
The top-quark mass value is one of key issues of the current researches.
%as it can be used to extrapolate the SM Higgs potential at energies up to the Plank scale. 
In particular, it plays an important role in testing the stability of the SM Higgs potential.
%Therefore, a precise top-quark mass measurement is required.
The most precise measurement of the top-quark mass is 172.44 $\pm$ 0.13 (stat.) $\pm$ 0.47 (syst.) GeV from the published CMS result~\cite{cms:mass}
at the LHC.
The result is limited by the systematic uncertainty, 
coming mainly from the modeling of the hadronization. 
The ATLAS collaboration has also measured the top-quark mass of 
172.84 $\pm$ 0.34 (stat.) $\pm$ 0.61 (syst.) GeV,
which is mostly limited by the systematic uncertainty from the jet energy scale~\cite{atlas:mass}.

These measured top-quark masses are believed to be different from the pole mass due to 
non-perturbative effects like hadronization.
To date, the pole mass has been measured with a relatively large uncertainty of around 2 GeV from the $\ttbar$ cross section measurements~\cite{cms:crossx-polemass,atlas:crossx-polemass} and using $\ttbar$+1 jet events~\cite{atlas:polemass}. 
In order to reduce the uncertainty, different approaches are to be explored.
One of such approaches is in the direction of utilizing leptonic observables~\cite{cdf:topmass,lepton:theory,topmasslep}.
Since leptons do not involve QCD activities, leptonic observables are advantageous to the extraction of the top-quark pole mass.
The latest measurement using lepton differential distributions~\cite{atlas:polemassdiff} has an uncertainty as close as 1.5 GeV. 

Among the approaches utilizing leptonic observables, the method proposed in Ref.~\cite{topmasslep} has an unique feature.
The observable required to this method is only the lepton energy distribution in the top-quark decay in the {\it laboratory} frame, 
while the theoretical prediction ``compared'' with it is just the lepton energy distribution in the top-quark {\it rest} frame.
In other words, this method has a boost-invariant nature and is independent of top-quark velocities.
This method is called ``weight function method'' as it uses a characteristic weight function $W(E_\ell,m)$.
The essence of the method is as follows: with the lepton energy distribution $D(E_\ell)$ from an experiment, there are infinite number of weight functions for which the following quantity,
\begin{equation}\label{equ:integral}
	I(m) = \int dE_\ell D(E_\ell) W(E_\ell,m),
\end{equation}
vanishes when the parameter $m$ is equal to the true mass value of the top quark, i.e. $I(m=m_t^{true})=0$.
For the detail of the method, see Refs.~\cite{sayaka:topmass,sayaka:weightfunction}, which originally proposed the weight function method for the mass reconstruction of the Higgs boson.

A practical defect of this method is the fact that it requires the whole energy distribution at parton level which includes the region outside the detector acceptance.
In the LHC experiments, events with low-energy leptons are not available due to a lepton $p_T$ trigger.
In Ref.~\cite{topmasslep}, this problem was coped with by compensating the low energy part of the distribution with Monte-Carlo (MC) events.
Since the compensating MC part has $m_t$ dependence, a way to independently extract the top-quark mass was invented.
In addition, in order to avoid uncertainties related to the MC events, a way of determining the normalization of the compensating part was devised.
Furthermore, an event selection was carefully chosen not to deform the parton-level distribution of the remaining part.
Although these devices work well, they induce additional complexity.

In this paper, we perform a feasibility study of the top-quark mass measurement using the weight function method at the LHC.
In contrast to the study in Ref.~\cite{topmasslep}, we use an unfolding technique to obtain the parton-level distribution, which would make this method simpler and  handier experimentally.
The lepton energy distribution at parton level is obtained by a simple bin-by-bin unfolding for the lepton+jets decay channel.
We adopt more realistic event selection and detector simulation than the study in Ref.~\cite{topmasslep}.
We also estimate some of the major uncertainties that are expected in realistic data analyses at the LHC.
We show that this method
can provide an independent verification of the top-quark mass measurement and provide pointers toward possible improvements with Run 2 data at the LHC.

\section{Samples}
\label{sample}

Simulated $pp$ collision data samples for the $\ttbar$ process are generated 
at a center-of-mass energy of 14 TeV, 
by using MadGraph5 (v2.4.0)~\cite{aMCNLO} at the leading order due to a limit of computer resources,
and interfaced with PYTHIA (v6.428)~\cite{pythia} for parton showering and hadronization.
The $\ttbar$ samples are generated for five different values of the top-quark mass, 
167, 170, 173, 176 and 179 GeV.
For each signal sample, 600K events are generated. 
A sample with a top-quark mass of 173 GeV is generated separately to obtain a detector response correction
required in the unfolding procedure. 
The sample for the unfolding is statistically independent of the signal data sample
with the size of 1200K to avoid any statistical bias.

To emulate a detector performance, the generated events are processed through the
DELPHES package (v3.3.2)~\cite{delphes} using the public CMS detector card.
Similar to the CMS reconstruction, the objects from the particle-flow algorithm implemented in DELPHES are used throughout
this analysis.

Pileup events are not simulated in this analysis.
Although the effects of pileup can be merged with the simulated events in the DELPHES package,
pileup mitigation, which will be developed
at the LHC experiments, can reduce the pileup effects significantly.
It is also important to understand the physics without pileup events.
Therefore, we focus on the physics under the condition that there are no pileup effects.

In the DELPHES fast simulation, momenta of all the physics objects
such as electrons, muons and jets, are smeared as a function of 
their transverse momenta ($\pt$) and pseudorapidities ($\eta$) 
so that the detector effects in the CMS experiment are simulated.
Reconstruction efficiencies of electrons, muons and jets are also parameterized as functions of $\pt$ and $\eta$
based on the public information from the CMS experiment.

The muon identification efficiency is set to 95\% for
$\pt$ $>$ 10 GeV and $|\eta|$ $<$ 2.4. 
The electron identification efficiency is set to 95\% for
$\pt$ $>$ 10 GeV and $|\eta|$ $<$ 1.5, and 85\% for $\pt$ $>$ 10 GeV and 1.5 $<$ $|\eta|$ $<$ 2.5.
Isolated muons and electrons are selected by applying a relative isolation of $I_{rel}$ $<$ 0.1, where $I_{rel}$
is defined as the sum of the surrounding energy of the particle-flow tracks, photons and neutral hadrons
divided by $\pt$ of the muon or electron.

The particle-flow jets used in this analysis are clustered by using particle-flow tracks and particle-flow towers.
If a jet is already reconstructed as an isolated electron, muon or photon, this jet is excluded from further consideration.
The $b$-tagging efficiency parameterized as a function of $\pt$ and $\eta$ 
ranges from 20\% to 50\%.
The fake $b$-tagging rate for light-flavor jets is set to 0.1\%,
which corresponds to the tight working point in the CMS paper Ref.~\cite{cmsbjet}.

\section{Event selection}{\label{selection}}

Events are selected based on the decay topology of the top-quark pair in the lepton+jets channel.
The event should have exactly one isolated lepton ($e, \mu$) with $\pt$ $>$ 20 GeV and $|\eta|$ $<$ 2.1.
Events are further selected by requiring at least four jets with $\pt$ $>$ 30 GeV and two $b$-tagged jets
to reject SM backgrounds such as $W$ + jets and single-top events.
The acceptance after all the requirements is 4.3\%.
After this typical event selection for the lepton+jets channel, 
the background contribution is expected to be less than 10\% level~\cite{lepjets}.
In the unfolding procedure which will be described in Section~\ref{analysis}, 
the remaining background after the selection
is assumed to be subtracted from data. 
A possible background contribution and its uncertainty are not considered in this analysis.

\section{Measurement}
\label{analysis}

\begin{figure}
\resizebox{0.5\textwidth}{!}{
\includegraphics{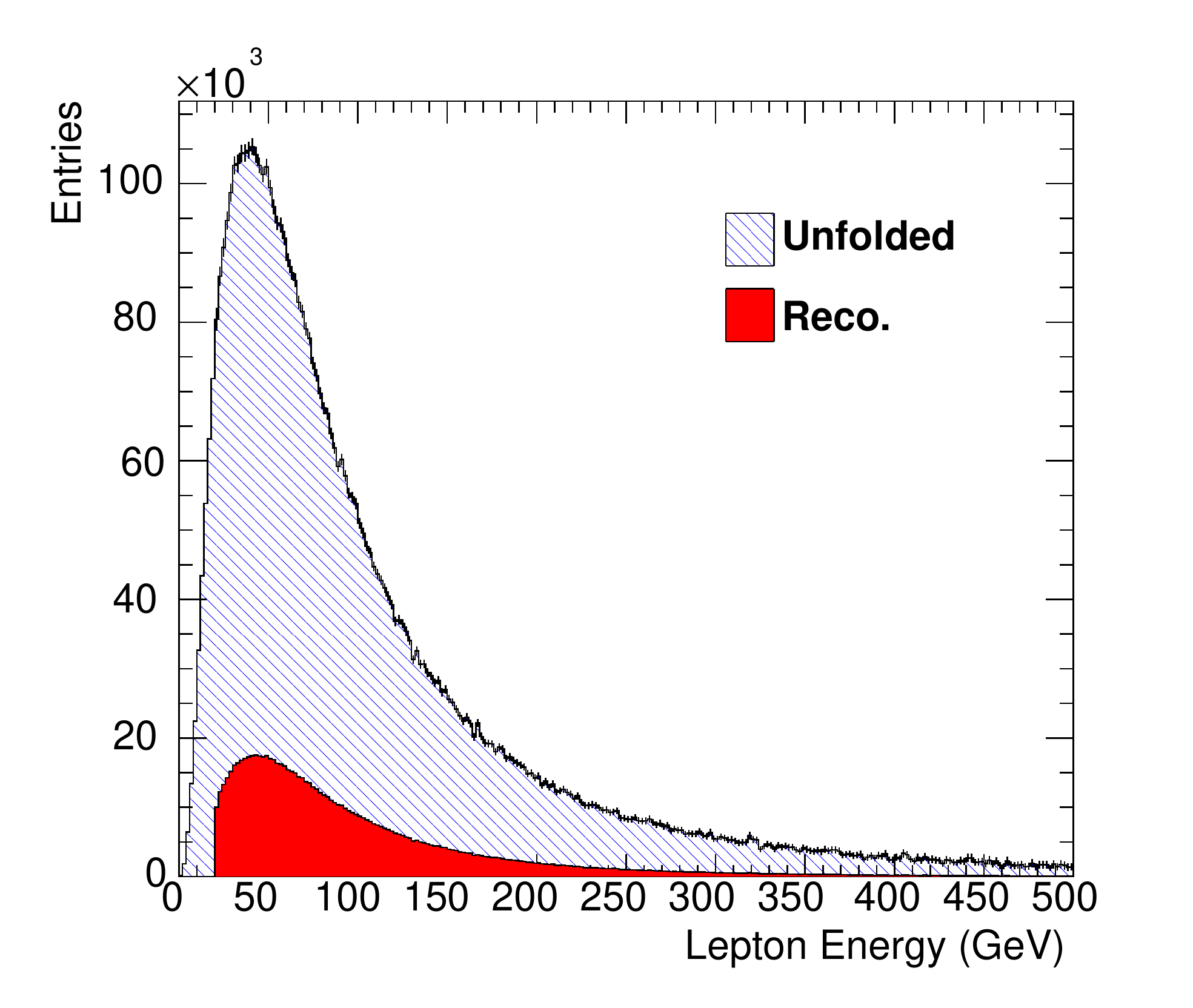}% Here is how to import EPS art
}
\caption{\label{fig:lep}
Energy distribution of lepton at reconstruction level for the top-quark mass of 173 GeV (red color) and the unfolded distribution at parton level (blue color). 
The energy distribution at reconstruction level is unfolded using an additional sample with the top quark mass of 173 GeV.
}
\end{figure}

The weight function method requires the lepton energy distribution at parton level.
In order to obtain the parton-level distribution,
we use an unfolding technique~\cite{roounfold} for removing effects of detector performance, photon radiation, lepton isolation, the event selection, etc. The lepton energy distribution at reconstruction level is unfolded back to the parton-level distribution
by using a simple bin-by-bin unfolding.
For the unfolding, the additional sample with the top-quark mass of 173 GeV is used.
The lepton energy distribution at reconstruction level and the unfolded distribution are shown in Fig.~\ref{fig:lep}.
A bin width of 2 GeV is used for the lepton energy distributions.
In reality, a more complicated unfolding such as regularization might be required to correct 
effects of detector resolution and bin migration,
which can arise from energy loss due to final-state radiation from a muon or Bremsstrahlung from an electon.

In this analysis, the unfolding is done in two steps.
The first step is to correct the event selection effect.
The energy distribution after the final selection 
is unfolded back to 
the distribution at preselection level with the single-lepton requirement.
In this first step, since events are within the acceptance range, 
a data-driven method could be used.
The second step is to correct detector effects such as acceptance and resolution on the lepton energy spectrum
in order to obtain the energy distribution at parton level
from the distribution at preselection level with the single-lepton requirement.
In this second step,
the response sample with the top-quark mass of 173 GeV is used.
In the unfolding procedure, it is important to have a statistically independent sample to avoid any bias.
Therefore, an additional 1200K events are generated for the response distribution. 

\begin{figure}[ht]
\resizebox{0.5\textwidth}{!}{
\includegraphics{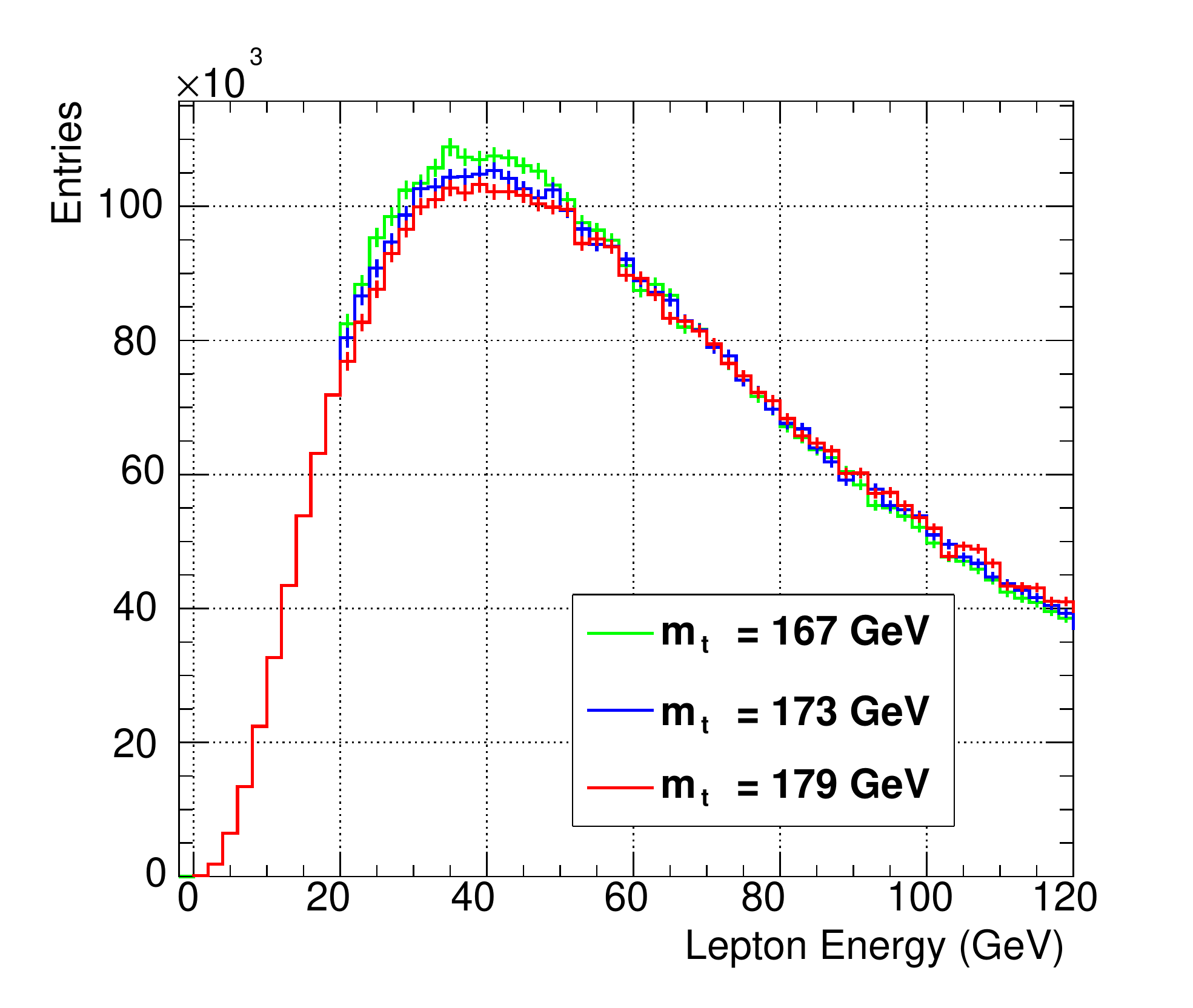}% Here is how to import EPS art
}
\caption{\label{fig:unfolded}
Unfolded energy distributions of lepton at parton level at the top-quark masses of 167, 173 and 179 GeV.
}
\end{figure}

At reconstruction level, there are no events in a low-energy region below 20 GeV
due to the lepton-trigger requirement of $\pt > 20$ GeV.
Therefore, we rely on the MC simulation below the $\pt$ threshold of 20 GeV.
%In real analysis, it is crucial to lower the $\pt$ threshold to have large acceptance
%to avoid the MC dependency in low lepton energy region where there are no data events available.
Figure~\ref{fig:unfolded} shows unfolded distributions at the top-quark masses of 167, 173 and 179 GeV.
The distribution below the 20 GeV threshold at parton level is from the response sample generated 
at $m_t$=173 GeV.

\begin{figure}[ht]
\resizebox{0.5\textwidth}{!}{
\includegraphics{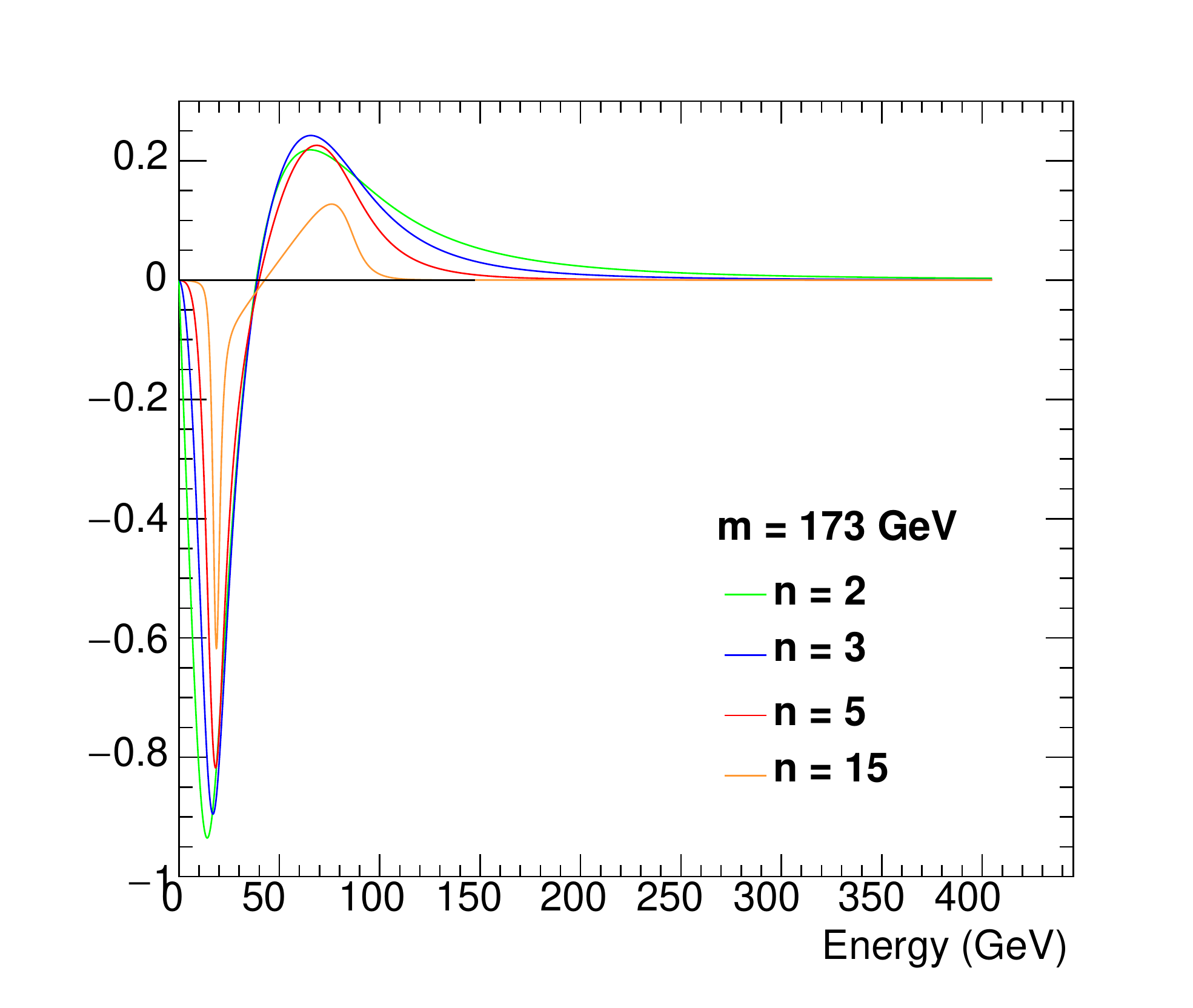}% Here is how to import EPS art
}
\caption{\label{fig:weight} Weight functions $W(E_\ell,m)$ provided by the authors of Ref.~\cite{topmasslep} corresponding to $n$ = 2, 3, 5, 15 in Eq.~(\ref{equ:weight}) with $m= 173$ GeV.}
\end{figure}

After the unfolding procedure,
the unfolded energy distribution at parton level is used in the weight function method.
The weight functions $W(E_\ell,m)$ used in this analysis are provided by authors of Ref.~\cite{topmasslep}.
The explicit form of the weight function is
\begin{equation}\label{equ:weight}
W(E_\ell,m) \propto \int dE \mathcal{D}_0 (E; m) \frac{1}{EE_\ell}\frac{(E_\ell/E)^n - (E/E_\ell)^n}{ [(E_\ell/E)^n + (E/E_\ell)^n]^2 }\,,
\end{equation}
where $\mathcal{D}_0(E;m)$ is a theoretical prediction at leading-order for the distribution of lepton energy $E$, calculated in the top-quark rest frame with a top-quark mass value $m$.
The weight functions corresponding to $n$ = 2, 3, 5, 15 for $m = 173$ GeV are shown in Fig.~\ref{fig:weight}.

With these weight functions, the following method is applied.
As explained in Section~\ref{sec:intro}, we calculate weighted integrals of the lepton energy distribution [Eq.~(\ref{equ:integral})].
With binned data, the integral in Eq.~(\ref{equ:integral}) is replaced by a sum:
\begin{equation}
	I(m) = \sum_i N_i W(E_i, m) / \sum_j N_j \,,
\end{equation}
where $N_i$ and $E_i$ are the number of entries and the lepton energy, respectively, for $i$-th bin of the lepton energy distribution.
Then the reconstructed top-quark mass $m_t^{rec}$ is extracted through $I(m=m_t^{rec})=0$.
We multiply the unfolded energy distribution and the weight function bin by bin, and obtain the weighted sum.

\begin{figure}[ht]
\resizebox{0.5\textwidth}{!}{
\includegraphics{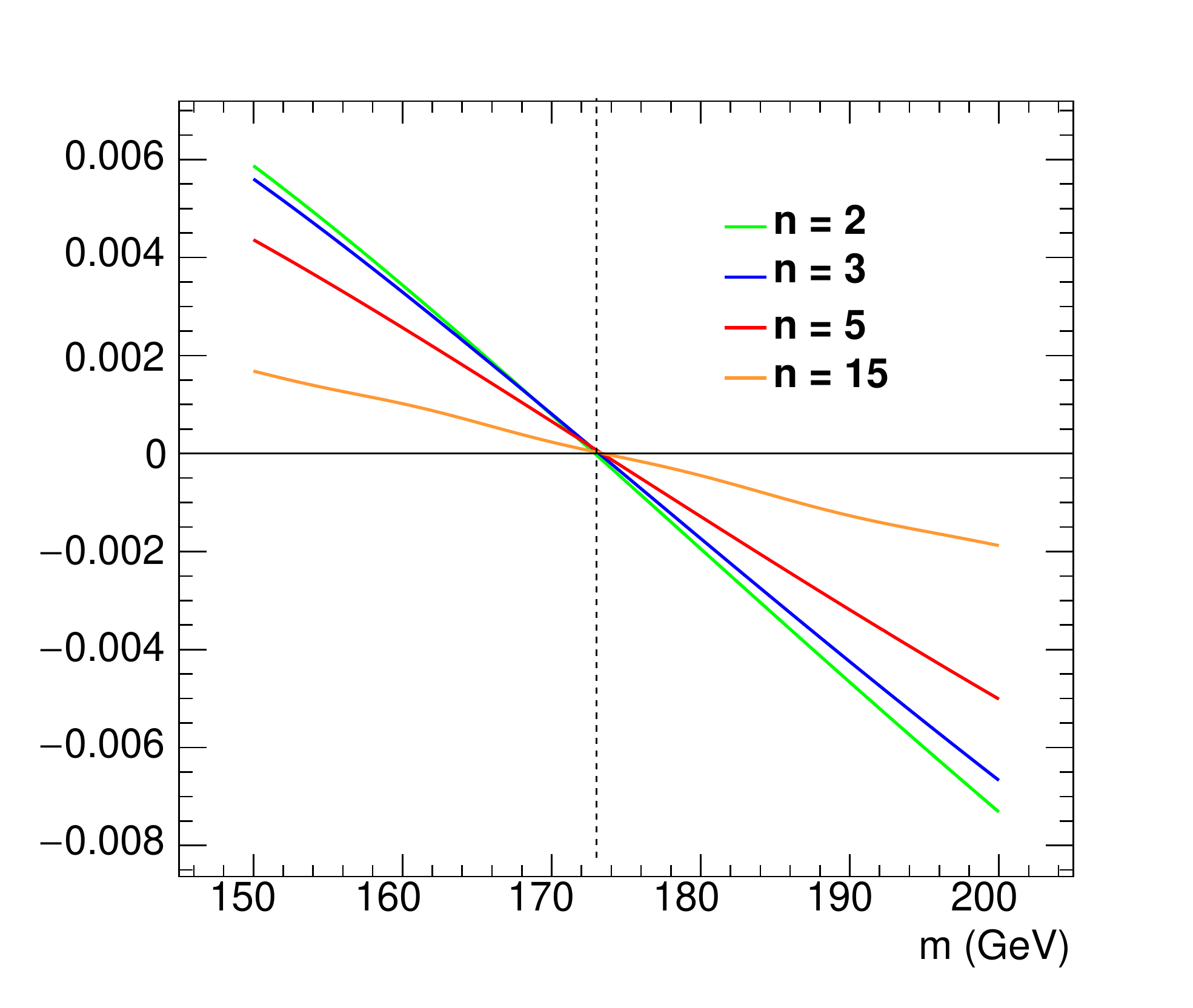}% Here is how to import EPS art
}
\caption{\label{fig:integral} Weighted sums $I(m)$ over the unfolded lepton energy distribution with the weight functions
corresponding to $n$ = 2, 3, 5, 15.
The input value of the top-quark mass is 173 GeV, as its position is indicated by a dashed line.
The horizontal axis is the mass parameter $m$ used 
in the calculation of $W(E_\ell,m)$.
The top-quark mass is reconstructed from $I(m=m_t^{rec})=0$.
}
\end{figure}

Figure~\ref{fig:integral} shows the weighted sums over the parton-level energy distribution with the weight functions corresponding to $n=$ 2, 3, 5, 15. The zeros of the weighted sums indicate
reconstructed top-quark masses.
In this figure, the input top-quark mass to the data sample is set to 173 GeV.
The plot shows that
the input value is correctly reconstructed for each $n$ using the unfolded energy distribution.

\section{Results}
\label{sec:result}

\begin{figure}[ht]
\resizebox{0.5\textwidth}{!}{
\includegraphics{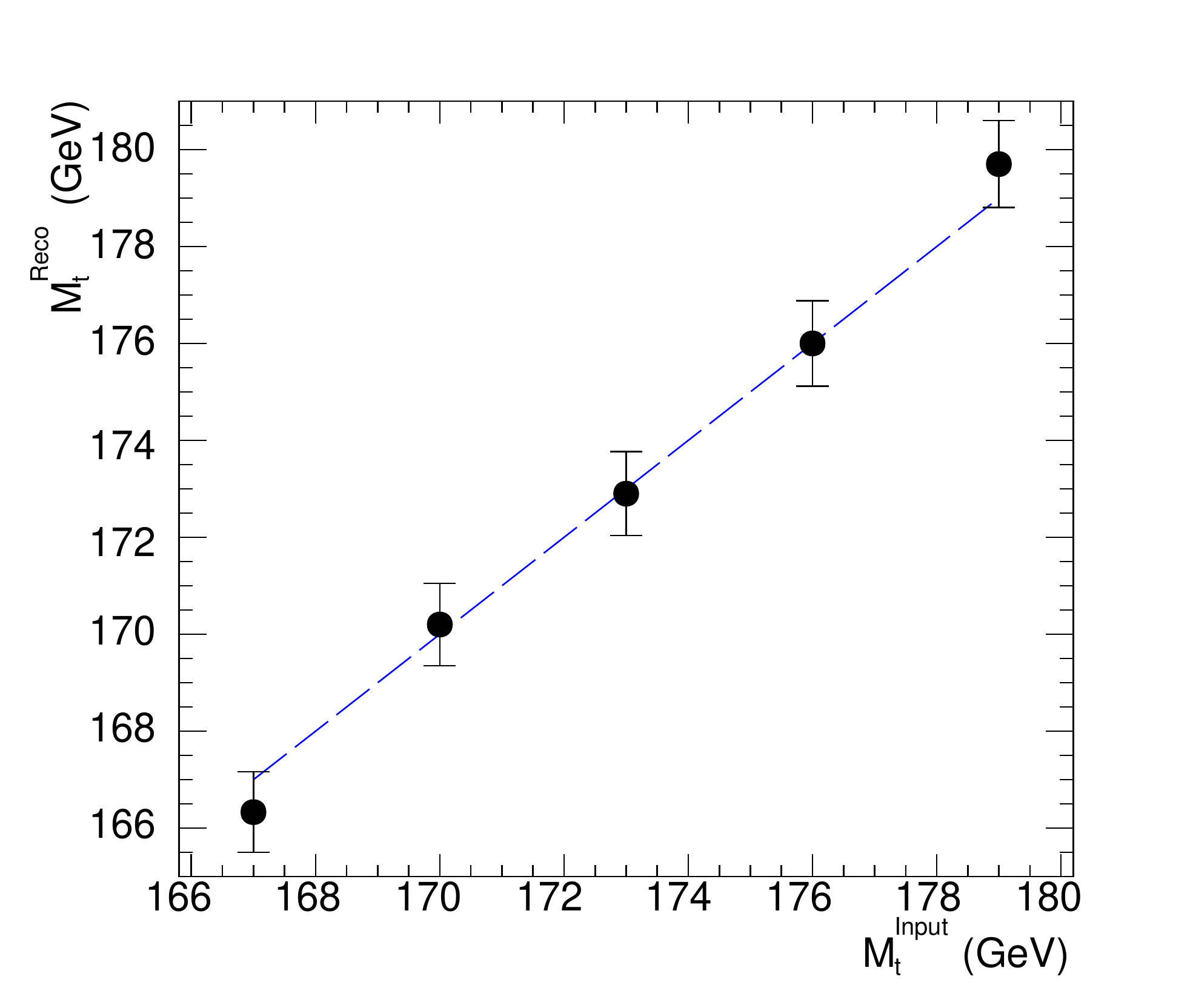}% Here is how to import EPS art
}
\caption{\label{fig:bias} Validation with different input values of the top-quark mass using the response distribution 
with $m_t = 173$ GeV. 
The vertical axis indicates the reconstructed top-quark mass and the horizontal axis indicates the input mass.
The dashed line indicates the ideal case where the input mass is the same as the reconstructed mass.
The statistical uncertainty is shown together with the central value.
}
\end{figure}

Figure~\ref{fig:bias} 
shows reconstructed top-quark masses versus input masses in a 3-GeV step for the $n=2$ weight function.
The response sample with the top-quark mass of 173 GeV is used for the unfolding including the correction 
below the 20 GeV threshold.
The statistical uncertainty is estimated by varying each bin randomly within the statistical uncertainty in the unfolded distribution.
The statistical uncertainty shown on the unfolded distribution is the sum of the statistical uncertainties from
the response distribution and the input distribution. This uncertainty on the unfolding
is obtained by running toy MC experiments.
The estimated statistical uncertainty is around 0.5\% which corresponds to about 0.8 GeV for the top-quark mass of 173 GeV.
The statistical uncertainty described above is shown in Fig.~\ref{fig:bias}.

Table~\ref{tab:table2} shows the input and reconstructed top-quark masses using the weight function 
corresponding to $n$=2, 3, 5, 15.
The reconstructed top-quark masses are consistent with the input top-quark masses within the sizes of the statistical error. 
Note that the results with different $n$ involve different sizes of statistical and systematic uncertainties, although they are strongly correlated.
%For larger n, there would be a systematic bias due to the 2 GeV bin width. 
%The effect increases for larger $n$ weight functions.  
In addition, there would be a systematic bias due to the 2-GeV bin width.
Since a weight function with larger $n$ has a sharper form in the low-energy region (see Fig. 3), 
the bias due to the bin width is larger for the larger-$n$ weight function.

\begin{table}[ht]%The best place to locate the table environment is directly after its first reference in text
\centering
\caption{\label{tab:table2}%
Input and reconstructed masses using the response distribution with $m_t = 173$ GeV.
The statistical uncertainty for each mass value is 0.5\%.
}
\begin{tabular}{c|ccccc}
\hline
input m$_t$ (GeV)   & 167    & 170   & 173   & 176   & 179    \\\hline
reco. m$_t$ (GeV)   &        &       &       &       &        \\
$n$ = 2               & 166.3  & 170.2 & 172.9 & 176.0 & 179.7  \\
$n$ = 3               & 166.4  & 170.4 & 173.2 & 176.3 & 179.9  \\
$n$ = 5               & 166.3  & 170.6 & 173.4 & 176.4 & 179.9  \\
$n$ = 15              & 166.6  & 170.8 & 173.4 & 176.7 & 179.5 \\

\hline
\end{tabular}
\end{table}

To see effects of the lepton $\pt$ threshold on the result,
the analysis is repeated with different lepton $\pt$ thresholds of 22, 24 and 26 GeV.
Increasing the threshold would lead to a larger bias on the reconstructed top-quark mass.
Table~\ref{tab:table3} shows input and reconstructed masses for various lepton $\pt$ 
thresholds. One can see in Table~\ref{tab:table3}
that the reconstructed top-quark mass approaches the top-quark mass value of the response sample,
namely 173 GeV, as the threshold increases. 
With the 20 GeV threshold, the bias is sufficiently small compared with the statistical uncertainty.

\begin{table}[ht]%The best place to locate the table environment is directly after its first reference in text
\centering
\caption{\label{tab:table3}%
Input and reconstructed masses
for various lepton $\pt$ thresholds from 20 GeV to 26 GeV in steps of 2 GeV.
The weight function for $n=2$ is used.
%The statistical uncertainty is 0.5\%.
}
\begin{tabular}{c|ccccc}
\hline
input m$_t$ (GeV)       & 167    & 170   & 173   & 176   & 179    \\\hline
reco. m$_t$ (GeV)       &        &       &       &       &        \\
$\pt > $ 20 GeV         & 166.3  & 170.2 & 172.9 & 176.0 & 179.7  \\
$\pt > $ 22 GeV         & 167.7  & 170.3 & 173.6 & 175.5 & 178.6  \\
$\pt > $ 24 GeV         & 168.7  & 170.9 & 173.6 & 175.3 & 177.4  \\
$\pt > $ 26 GeV         & 169.8  & 171.1 & 173.3 & 174.9 & 176.2  \\
\hline
\end{tabular}
\end{table}

\section{Discussion}
\label{sec:disc}
In this section, we discuss and estimate main systematic uncertainties that can arise in this method.
The most serious bias is caused by the fact we have to rely on the response distribution from the MC sample 
for the lepton energy distribution below the threshold.
As the lepton energy threshold goes up, we rely more and more on the response sample.
%which has usually larger statistics than the input distribution.
%There would be trade off between statistical uncertainty and systematic uncertainty in increasing the threshold.
In Section~\ref{sec:result}, we showed this possible bias is negligible when the lepton energy threshold of 20 GeV is used.
The result with this threshold is consistent with the input top-quark mass within the statistical uncertainty of 0.5\%.
However, with a threshold of above 20 GeV, the bias becomes larger than the statistical uncertainty, and
the method used in this paper has a difficulty.
Therefore, it is desired that the lepton $\pt$ threshold is as low as possible.
Note that this is not a critical problem for the weight function method itself. 
As explained in Section~\ref{sec:intro}, this problem is solved in Ref.~\cite{topmasslep}, 
by imposing a consistency condition that the reconstructed mass is equal to the mass for the MC sample below the threshold. 
The result in our study demonstrates that the simple way used in this analysis can be applied only with a $\pt$ trigger 
as low as 20 GeV. With a lepton $\pt$ trigger above 20 GeV, 
it would be essential to apply a supplementary method like the one described in Ref.~\cite{topmasslep}.
%Because it would be challenging to keep the lepton $\pt$ threshold this low in experiments in Run 2 at the LHC,
%additional treatments would be required in a real analysis, for example, a calibration of the obtained mass and a check for an appropriate
%m$_{t}$ value in the response distribution by using other distributions.

There would be a systematic uncertainty from the data-driven correction 
in the first step of the unfolding described in Section~\ref{analysis}. 
In order to obtain the correction factor for the selection efficiency, 
an orthogonal event selection can be applied to enhance top-quark events. 
However, it would not be trivial to find such an orthogonal event selection.
In this analysis, the uncertainty in this step is ignored but can be significant in a real analysis.   
In addition, uncertainties on the background subtraction are not included in this analysis but can contribute.
The largest experimental uncertainty in the top-quark pair cross section used to determine the pole mass 
is from the lepton energy scale and resolution~\cite{cms:crossx-polemass,atlas:crossx-polemass}.
%The uncertainty is assessed as 1\% level in ATLAS and CMS experiments~\cite{atlas:crossx-polemass,cms:crossx-polemass}.

Uncertainties from the factorization and renormalization scales are estimated by varying the scales for input distributions
by a factor of two up and down with respect to their reference values for the lepton energy distribution at the top-quark mass of 173 GeV.
The same response sample with the nominal scale is kept for the unfolding.
The uncertainty of 0.3\% is assessed by taking the difference in the result.
It should be noted that these scale uncertainties are the dominant ones in the estimate of uncertainties 
in the study of Ref.~\cite{topmasslep}. 
The smallness of the scale uncertainties compared to those in Ref.~\cite{topmasslep} is an advantage of using the unfolding.
It would also be required to have an extensive validation of the unfolding by using different MC generators to check any possible bias from theoretical predictions.
In particular, the validation with NLO generators would be very useful. 
%Additional uncertainty would come from the different theoretical predictions in matrix element (ME).

The weight function method is based on the assumption that the top quark is on-shell.
Thus, the actual finite width of the top quark causes a deviation to the reconstructed mass.
We estimate the size of this deviation by examining the invariant mass distribution of the
top quark at parton level. With the parameter setting that the cutoff for
the Breit-Wigner distribution in the configuration of the MadGraph package is at
$m_t\pm 50 \Gamma_t$, the mean value of the mass distribution is shifted from
the input mass value by the amount of 0.3 GeV for each mass. Therefore, we expect a
systematic shift of the order of 0.3 GeV in the result for the reconstructed mass.
In real experiments, the effect of the top-quark width can be estimated by
simulation analyses with MC generators which take into account the top-quark finite-width effects more thoroughly (see, for example, Ref.~\cite{topquarkwidth}).
%The final result for the reconstructed mass is obtained after a subtraction of the systematic 
%shift due to the width effects.

Overall, the sum of the statistical and systematic uncertainties 
considered in this analysis is less than 1 GeV.

This study is performed at leading order,
using the leading-order event generator and the leading-order theoretical prediction
for the lepton energy distribution $\mathcal{D}_0$.
If we include the next-to-leading order (NLO) corrections to them in the on-shell scheme,
a reconstructed mass is identified with the top-quark pole mass. More specifically, a weight
function at NLO is calculated with $\mathcal{D}^{\rm NLO}_0(E;m^{\rm pole})$, which is
the NLO distribution of lepton energy (in the top-quark rest frame) with a top-quark pole mass
$m^{\rm pole}$. Then one can read off the NLO pole mass value from a weighted integral
through  $I(m^{\rm pole}=m_t^{{\rm pole},\,rec})=0$. Note that for the purpose of extracting the NLO
pole mass, only the corrections to the top-quark decay process are required 
since the theoretical prediction used in the weight functions, $\mathcal{D}_0^{\rm (N)LO}$, is the distribution in the top-quark rest frame. 
Thanks to the boost-invariant nature of the weight function method (as mentioned in Section~\ref{sec:intro} and 
proven in Refs.~\cite{topmasslep,sayaka:topmass}), the method relies on only the top-quark decay process for theoretical prediction. 
Note also that
the theoretical corrections to the top-quark decay process are much smaller than those to
the production process, so that experimental consequences are insensitive to the corrections to
the decay process. These suggest that the application of this analysis to the NLO pole-mass extraction
is straightforward and the experimental uncertainties estimated in this leading-order study
will not change significantly.

\section{Conclusions}
We estimated the sensitivity of the top-quark mass measurement with the weight function method 
by using simulation samples.
This method requires only the lepton energy distribution at parton level.
Events with exclusively one lepton, at least four jets and two $b$ jets are selected.
The lepton energy distribution at reconstruction level after the selection
is unfolded back to the energy distribution at parton level.
In the region below the energy threshold of 20 GeV, the response sample with the top-quark mass of 173 GeV is used. 
The reconstructed mass of the top quark with the weight function method is consistent with the input mass 
within the statistical uncertainty of 0.8 GeV.
%There is no significant bias from the MC sample below the energy threshold of 20 GeV.
We discussed and estimated some of main systematic uncertainties expected in this method. 
Taking into account the statistical and systematic uncertainties,
the estimated uncertainty of the reconstructed top-quark mass is of the order of 1 GeV. 
This uncertainty is compatible with the current uncertainty of 2 GeV in the measurement of 
the top-quark pole mass.
Therefore, the weight function method could provide an alternative approach to measure the top-quark mass
without introducing large systematic uncertainties that can arise due to the jet energy measurement.
This study shows that the weight function method can also provide an independent verification of the top-quark mass measurement and provide pointers toward possible improvements with Run 2 data.

%\begin{acknowledgements}
\section*{ACKNOWLEDGMENTS}
This work was supported by the research fund of Hanyang University (HY-2015).
This work was supported by a National Research Foundation of Korea Grant
funded by the Korean governments (NRF-2014R1A1A2056283 and NRF-2014S1A2A2028546).
The work of S.K. was supported by Basic Science Research Program through 
the National Research Foundation of Korea (NRF) funded by the Ministry of Science, ICT and Future Planning (Grant No. NRF-2014R1A2A1A11052687).
%\end{acknowledgements}


\begin{thebibliography}{99}

\bibitem{cms:mass} CMS Collaboration, \textit{Measurement of the top quark mass using proton-proton collisions at $\sqrt{s}$ = 7 and 8 TeV}, Phys. Rev. D \textbf{93} (2016) 26, doi:10.1103/PhysRevD.93.072004, arXiv:1509.04044.

\bibitem{atlas:mass} ATLAS Collaboration, \textit{Measurement of the top quark mass in the $t\bar{t} \to$ dilepton channel from $\sqrt{s}$ = 8 TeV ATLAS data}, Phys. Lett. B \textbf{761}, 350 (2016), 10.1016/j.physletb.2016.08.042, arXiv:1606.02179.  

\bibitem{cms:crossx-polemass} CMS Collaboration, \textit{Measurement of the $t\bar{t}$ production cross section in the $e\mu$ channel in proton-proton collisions at $\sqrt(s)$ = 7 and 8 TeV}, JHEP \textbf{08} (2016) 029, 10.1007/JHEP08(2016)029, arXiv:1603.02303.

\bibitem{atlas:crossx-polemass} ATLAS Collaboration, \textit{Measurement of the $t\bar{t}$ productio n cross section in the $e\mu$ events with b-tagged jets in pp collisions at $\sqrt(s)$ = 7 and 8 TeV with the ATLAS detector}, Eur. Phys. J. C \textbf{74} (2014) 3109, 10.1140/epjc/s10052-014-3109-7, arXiv:1406.5375.

\bibitem{atlas:polemass} ATLAS Collaboration, \textit{Determination of the top-quark pole mass using $t\bar{t}$ + 1-jet events collected with the ATLAS experiment in 7 TeV pp collisions}, arXiv:1507.01769.

\bibitem{cdf:topmass} CDF Collaboration, \textit{Measurement of the Top Quark Mass in the Lepton+Jets Channel Using the Lepton Transverse Momentum}, Phys. Lett. B \textbf{698}, 371 (2011), 10.1016/j.physletb.2011.03.041, arXiv:1101.4926.

\bibitem{lepton:theory} S. Frixione and A. Mitov, \textit{Determination of the top quark mass from leptonic observables}, JHEP \textbf{09} (2014) 012, doi:10.1007/JHEP09(2014)012, arXiv:1407.2763.

\bibitem{topmasslep} S. Kawabata, Y. Shimizu, Y. Sumino and H. Yokoya, \textit{Weight function method for precise determination of top quark mass at Large Hadron Collider}, Phys. Lett. B \textbf{741}, 232 (2014), 10.1016/j.physletb.2014.12.044, arXiv:1405.2395.

\bibitem{atlas:polemassdiff} ATLAS Collaboration, \textit{Measurement of lepton differential distributions and the top quark mass in $t\bar{t}$ production in pp collisions at $\sqrt{s}$ = 8 TeV with the ATLAS detector}, arXiv:1709.09407.

\bibitem{sayaka:topmass} S. Kawabata, Y. Shimizu, Y. Sumino and H. Yokoya, \textit{Boost-Invariant Leptonic Observables and Reconstruction of Parent Particle Mass}, Phys. Lett. B \textbf{710} (2012) 658-664, 10.1016/j.physletb.2012.03.050, arXiv:1107.4460. 

\bibitem{sayaka:weightfunction} S. Kawabata, Y. Shimizu, Y. Sumino and H. Yokoya, \textit{Measurement of physical parameters with a weight function method and its application to the Higgs boson mass reconstruction}, JHEP \textbf{08} (2013) 129, 10.1007/JHEP08(2013)129, arXiv:1305.6150.

\bibitem{aMCNLO} J. Alwall et al.,
%R. Frederix, S. Frixione, V. Hirschi, F. Maltoni, O. Mattelaer, H.-S. Shao, T. Stelzer, P. Torrielli, and M. Zaro, 
\textit{The automated computation of tree-level and next-to-leading order differential cross sections, and their matching to parton shower simulations}, JHEP \textbf{07} (2014) 079, doi:10.1007/JHEP07(2014)079, arXiv:1405.0301.

\bibitem{pythia} T. Sj{\"o}strand, S. Mrenna, and P. Skands, 
\textit{PYTHIA 6.4 physics and manual}, JHEP \textbf{05} (2006) 026, 10.1088/1126-6708/2006/05/026, arXiv:hep-ph/0603175.

\bibitem{delphes} J. de Favereau et al.,
%C. Delaere, P. Demin, A. Giammanco, V. Lemaître, A. Mertens, and M. Selvaggi,
\textit{DELPHES 3, A modular framework for fast simulation of a generic collider experiment} (2013), JHEP \textbf{02}, 057 (2014), 10.1007/JHEP02(2014)057, arXiv:1307.6346.

\bibitem{cmsbjet} CMS Collaboration, \textit{Identification of b-quark jets with the CMS experiment}, J. Instrum. \textbf{8}, P04013 (2013), 10.1088/1748-0221/8/04/P04013, arXiv:1211.4462.

\bibitem{lepjets} CMS Collaboration, \textit{Measurement of differential $t\bar{t}$ production cross sections in lepton + jets final states at 13 TeV}, arXiv:1610.04191.

\bibitem{roounfold} Tim Adye, \textit{Unfolding algorithms and tests using RooUnfold} (2011), arXiv:1105.1160.

\bibitem{topquarkwidth} T. Jezo, J.M. Lindert, P. Nason, C. Oleari and S. Pozzorini, \textit{An NLO+PS generator for $t\bar{t}$ and $Wt$ production and decay including non-resonant and interference effects}, Eur. Phys. J. C \textbf{76} (2016) 691, 10.1140/epjc/s10052-016-4538-2, arXiv:1607.04538. 

\end{thebibliography}
\end{document}